\documentstyle[12pt,thmsa,sw20lart]{article}

\input{tcilatex}

\begin{document}

\title{Exact renormalization group equations and the field theoretical approach to critical phenomena}

\author{C. Bagnuls\thanks{%
Service de Physique de l'Etat Condens\'{e} e-mail: bagnuls@spec.saclay.cea.fr%
} \ and C.\ Bervillier\thanks{%
Service de Physique Th\'{e}orique e-mail: bervil@spht.saclay.cea.fr} \\
C. E. Saclay, F91191 Gif-sur-Yvette Cedex, France}
\maketitle

\begin{abstract}
After a brief presentation of the exact renormalization group equation, we
illustrate how the field theoretical (perturbative) approach to critical
phenomena takes place in the more general Wilson (nonperturbative) approach.
Notions such as the continuum limit and the renormalizability and the
presence of singularities in the perturbative series are discussed.
\end{abstract}

This paper has two parts. In the first part we restrict ourselves to the
presentation of some selected issues taken from a recent review on the exact
renormalization group (RG) equation (ERGE) in the pure scalar case.~\cite
{4595} In the second part we illustrate the Wilson continuum limit of field
theory (or, equivalently, what may be understood as the nonperturbative
renormalizability). More generally we would like to indicate how the
historical first version of the RG, based on a perturbative approach, takes
place in the more general nonperturbative framework developed by Wilson. The
illustration will be done in the light of actual RG trajectories obtained
from a numerical study of the ERGE in the local potential approximation.~%
\cite{3554}

\section{Selected issues\label{sec: I}}

\subsection{What does ``exact'' mean ?}

The word ``exact'' means: a continuous (i.e. not discrete) realization of
the Wilson RG transformation of the action in which no approximation is made
and no expansion is involved with respect to some small parameter. We are
here only interested in the differential form of the ERGE. One could think
that the word exact is not adapted to the Wilson renormalization because,
compared to the standard version of renormalization\footnote{%
Which originates from perturbation theory.}, it involves a finite cutoff and
it is only a semi-group. Nevertheless the historical first version is
entirely contained in the Wilson theory (see part \ref{sec: II}).

\underline{Vocabulary}: With a view to simplify the expression, in the following, the word ``{\em %
action}'' will systematically replace the customary expression ``{\em %
Wilson's effective action}'' while the expression ``{\em effective action}''
will be systematically used instead of ``{\em Legendre effective action}''
or ``{\em average effective action}''.

\subsection{Scheme dependence}

In addition to its complexity [an integro-differential equation, see eq. (%
\ref{WERGE})], there is not a unique form of the ERGE. Each form of the
equation is characterized by the way the momentum cutoff $\Lambda _{0}$ is
introduced. One says that the ERGE is scheme dependent\footnote{%
In perturbative RG, this kind of dependence is known as the {\em %
regularization}-scheme dependence or ``scheme dependence'' in short. This
should not be confused with the {\em renormalization} scheme dependence, a
notion associated with the continuum limit which embodies an arbitrariness
(see part \ref{sec: II}).}, but its various forms embody a unique physical
content in the sense that they all preserve the same physics at large
distances and, via the recourse to a process of limit, yield the same
physics at small distances (continuum limits). There exist some studies on
the scheme ``{\em independence''}.~\cite{SchemeDep}

Because a sharp boundary in momentum space introduces non-local interactions
in position space,~\cite{440} the first version of the ERGE has been
presented as far back as 1970~\cite{4495} with a smooth cutoff introduced
via an ``incomplete'' integration in which large momenta are more completely
integrated than small momenta\footnote{%
An other equation has been derived by Wegner and Houghton~\cite{414} for a
sharp cutoff and a third one by Polchinski~\cite{354} for field theoretical
purposes.}. The Wilson version of the ERGE reads:~\cite{440}

\begin{equation}
\dot{S}={\cal G}_{{dil}}S+\int_{p}\left( c+2p^{2}\right) \left( \frac{%
\delta ^{2}S}{\delta \phi _{p}\delta \phi _{-p}}-\frac{\delta S}{\delta \phi
_{p}}\frac{\delta S}{\delta \phi _{-p}}+\phi _{p}\frac{\delta S}{\delta \phi
_{p}}\right)  \label{WERGE}
\end{equation}
in which $S\left[ \phi \right] $ is the most general action\footnote{%
Generically speaking: compatible with the symmetry properties of the problem
studied.} that one may imagine for a scalar field, $\dot{S}=\frac{{d}S}{%
{d}t}$ (with $t=-\ln \frac{\Lambda }{\Lambda _{0}}$). The second term in
(\ref{WERGE}) corresponds to the reduction of the degrees of freedom
associated with the integration of the high frequency modes of the field and 
${\cal G}_{{dil}}S$ corresponds to the rescaling step of the RG
transformation:

\begin{equation}
{\cal G}_{{dil}}S=-\left( \int_{p}\phi _{p}\,{\bf p}\cdot \partial _{p}%
\frac{\delta }{\delta \phi _{p}}+d_{\phi }\int_{p}\phi _{p}\frac{\delta }{%
\delta \phi _{p}}\right) S  \label{Gdil}
\end{equation}
in which $d_{\phi }$ is the classical dimension of the field\footnote{%
In principle $d_{\phi }=\frac{d-2}{2}$, but because of the necessity of
adjusting the parameter $c$ to get a useful fixed point ---where the
anomalous dimension $\eta $ of the field may be non-zero--- it is customary
to let $d_{\phi }$ adjustable instead of $c$, in which case one writes $%
d_{\phi }=\frac{d-2+\eta }{2}$ with $\eta $ an adjustable constant (near a
fixed point).\label{footeta}}:

\begin{equation}
\phi _{sp}=s^{d_{\phi }-d}\phi _{p}  \label{dimfeffFou}
\end{equation}

The function $c(t)$ must be adjusted in such a way as to obtain a useful
fixed point.~\cite{440,4420}

\subsection{Reparametrization invariance}

This adjustment is a consequence of the independence~\cite{4420} of the RG
transformation on the overall normalization of $\phi $ which is also called
``reparametrization invariance''.~\cite{3357} This invariance occurs when
the RG transformation is such that the transformed field is related to the
original (unintegrated) field by a constant factor as it is presently the
case. As consequences:

\begin{itemize}
\item  a line of equivalent fixed points exists which is parametrized by the
normalization of the field,

\item  a field-rescaling parameter (it is often $\eta $ with $d_{\phi }=%
\frac{1}{2}\left( d-2+\eta \right) $, see footnote \ref{footeta}) takes on
a specific value.

\item  A redundant operator, associated to the change of normalization:
\end{itemize}

\[
{\cal O}_{1}=\int_{q}\left[ \frac{\delta ^{2}S}{\delta \phi _{q}\delta \phi
_{-q}}-\frac{\delta S}{\delta \phi _{q}}\frac{\delta S}{\delta \phi _{-q}}%
+\phi _{q}\frac{\delta S}{\delta \phi _{q}}\right] 
\]
has the eigenvalue~\cite{4405} $\lambda _{1}=0$ and is absolutely marginal%
\footnote{%
This is an exception due to the invariance otherwise, in general, a redundant
operator does not have a well defined eigenvalue.~\cite{2835}}.

\subsection{Effective action}

One may also consider an ERGE for the effective action\footnote{%
The effective action is the generating functional of the one-particle
irreducible vertex functions.} $\Gamma [\Phi ]$. It essentially allows to
circumvent the singularities induced by the hard cutoff and transforms the
UV cutoff into an IR cutoff.

We consider the action with an ``additive'' IR cutoff $\Lambda $ such that:~%
\cite{4374,3357}

\begin{equation}
{S}_{\Lambda }{[\phi ]\equiv }\frac{1}{2}\int_{p}\phi _{p}\phi
_{-p}C^{-1}(p,\Lambda )+S_{\Lambda _{0}}[\phi ]  \label{SlambdaIR}
\end{equation}
in which $C(p,\Lambda )$ is an additive infrared cutoff function which is
small for $p<\Lambda $ and $p^{2}C(p,\Lambda )$ should be large for $%
p>\Lambda $.

The ERGE may then be written as follows:~\cite{4315} 
\begin{equation}
\dot{\Gamma}={\cal G}_{{dil}}\Gamma +\frac{1}{2}{tr}\tilde{\partial%
}_{t}\ln \left( C^{-1}+\frac{\delta ^{2}\Gamma }{\delta \Phi \delta \Phi }%
\right)  \label{JungWet}
\end{equation}
in which $\tilde{\partial}_{t}\equiv -\Lambda \frac{\partial }{\partial
\Lambda }$ acts only on $C$ and not on $\Gamma $, i.e. $\tilde{\partial}%
_{t}=\left( \partial C^{-1}/\partial t\right) \left( \partial /\partial
C^{-1}\right) $.

\underline{A field theorist's self-consistent approach}: There is an efficient short cut for obtaining the ERGE satisfied by the
effective action. It is based on the observation that (\ref{JungWet}) may be
obtained from the one loop (unregularized, thus formal) expression of the
effective action, which reads (up to a field independent term within the
logarithm): 
\begin{equation}
\Gamma \left[ \Phi \right] =S\left[ \Phi \right] +\frac{1}{2}{tr}\ln
\left( \left. \frac{\delta ^{2}S}{\delta \phi \delta \phi }\right| _{\phi
=\Phi }\right) +{higher loop-order,}  \label{loop1}
\end{equation}
by using the following practical rules:

\begin{enumerate}
\item  add the infrared cutoff function $C(p,\Lambda )$ of eq. (\ref
{SlambdaIR}) within the action $S$, eq. (\ref{loop1}) then becomes: 
\[
\Gamma \left[ \Phi \right] =\frac{1}{2}\int_{p}\Phi _{p}\Phi
_{-p}C^{-1}(p,\Lambda )+S\left[ \Phi \right] +\frac{1}{2}{tr}\ln \left.
\left( C^{-1}+\frac{\delta ^{2}S}{\delta \phi \delta \phi }\right) \right|
_{\phi =\Phi }+\cdots 
\]

\item  redefine $\tilde{\Gamma}\left[ \Phi \right] =\Gamma \left[ \Phi
\right] -\frac{1}{2}\int_{p}\Phi _{p}\Phi _{-p}C^{-1}(p,\Lambda )$, then: 
\[
\tilde{\Gamma}\left[ \Phi \right] =S\left[ \Phi \right] +\frac{1}{2}{tr}%
\ln \left. \left( C^{-1}+\frac{\delta ^{2}S}{\delta \phi \delta \phi }%
\right) \right| _{\phi =\Phi }+\cdots 
\]

\item  perform the derivative with respect to $\Lambda $, (only the cutoff
function is concerned) and forget about the higher loop contributions: 
\[
\partial _{t}\tilde{\Gamma}={\frac{1}{2}}{tr}\left[ {\frac{1}{C}}%
\Lambda {\frac{\partial C}{\partial \Lambda }}\cdot \left( 1+C\cdot \frac{%
\delta ^{2}S[\Phi ]}{\delta \Phi \delta \Phi }\right) ^{-1}\right] 
\]

\item  replace\footnote{%
This step is often referred to as the ``{\em renormalization group
improvement''} of the one loop effective action.} the action $S$, in the
right hand side of the latter equation, by the effective action $\tilde{%
\Gamma}$ to get (\ref{JungWet}), the dilatation part ${\cal G}_{{dil}}%
\tilde{\Gamma}$ being obtained from usual (engineering) dimensional
considerations\footnote{%
But do not forget to introduce the anomalous dimension $\eta $ in order to
get an eventual nontrivial fixed point.}.
\end{enumerate}

It is noteworthy that the above rules have been heuristically first used~%
\cite{4430} to obtain the local potential approximation of the ERGE for the
effective action. However the main interest of the above considerations is
that they allow introducing the (infra-red) cutoff function independently of 
$S$, via the so-called ``{\em proper time}'' (or ``{\em heat kernel}'' or ``%
{\em operator}'') regularization.~\cite{4878} This kind of regularization is
introduced at the level of eq. (\ref{loop1}) via the general identity: 
\[
{tr}\ln \left( \frac{A}{B}\right) =-\int_{0}^{\infty }\frac{{d}s}{s%
}{tr}\left( {e}^{-sA}-{e}^{-sB}\right) 
\]

Forgetting again about the field-independent part (and, momentaneously,
about the ultra-violet regularization needed for $s\rightarrow 0$) one
introduces an infrared cutoff function $F_{\Lambda }\left( s\right) $ within
the proper time integral representation of the logarithm of $A=\left. \frac{%
\delta ^{2}S}{\delta \phi \delta \phi }\right| _{\phi =\Phi }$: 
\[
\frac{1}{2}{tr}\ln A\longrightarrow -\frac{1}{2}\int_{0}^{\infty }\frac{%
{d}s}{s}F_{\Lambda }\left( s\right) {tr\ e}^{-sA} 
\]

The function $F_{\Lambda }\left( s\right) $ must tend to zero sufficiently
rapidly for large values of $s$ in order to suppress the small momentum
modes and should be equal to 1 for $\Lambda =0$.

Then following the rules 3-4 above applied on $\Gamma $ (i.e., not on $%
\tilde{\Gamma}$), one obtains a new kind of ERGE\footnote{%
Notice that, because one performs a derivative with respect to $\Lambda $,
the essential contribution to $\partial _{t}\Gamma $ comes from the
integration over a small range of values of $s$ (corresponding to the rapid
decreasing of $F_{\Lambda }\left( s\right) $), hence an ultraviolet
regularization is not needed provided that the resulting RG equation be
finite.} for the effective action:~\cite{4858} 
\begin{equation}
\partial _{t}\Gamma =-\frac{1}{2}\int_{0}^{\infty }\frac{{d}s}{s}%
\Lambda \frac{\partial F_{\Lambda }\left( s\right) }{\partial \Lambda }\exp
\left[ -s\frac{\delta ^{2}\Gamma }{\delta \Phi \delta \Phi }\right]
\label{properge}
\end{equation}

There are apparently two advantages of using this kind of ERGE:

\begin{itemize}
\item  the regularization preserves the symmetry of the action,~\cite{4877}

\item  the derivative expansion is slightly easier to perform than in the
conventional approach and one may preserve the reparametrization invariance.~%
\cite{4858}
\end{itemize}

Bonano and Zappal\`{a}~\cite{4858} have considered the next-to-leading order
of the derivative expansion of (\ref{properge}) (while in a previous work~%
\cite{4873} only a pseudo derivative expansion, in which the wave-function
renormalization function $Z(\phi ,t)$ is field-independent, was used). They
have chosen $F_{\Lambda }\left( s\right) $ in such a way that the
integro-differential character of the ERGE disappears and they have tested
the preservation of the reparametrization invariance. Moreover a scheme
dependence parameter, related to the cutoff width, is at hand in this
framework which, presumably, will allow someone to look at the best possible
convergence of the derivative expansion when higher orders will be
considered.

Another kind of regularization related to this ``self-consistent'' approach
should be mentioned here. It consists in introducing the cutoff function in (%
\ref{loop1}) in-between the momentum integration [expressing the trace] and
the logarithm. This procedure has been considered at the level of the local
potential approximation.~\cite{4857} However the ERGE keeps its
integro-differential character and the study~\cite{4857} has then been
conducted within a constraining polynomial expansion method.

\subsection{Lowest order of the derivative expansion}

The study of the ERGE requires the use of approximation (and/or truncation)
methods such as the derivative expansion.~\cite{4468} It is a functional
power series expansion of the action in powers of momenta so that all powers
of the field are included at each level of the approximation.

In the position space it follows: 
\[
{S[\phi ]}={\int \!d^{d}x\,}\left\{ {V(\phi ,t)+}\frac{1}{2}{Z(\phi
,t)(\partial _{\mu }\phi )^{2}+\cdots }\right\} 
\]

The first order of the derivative expansion is the local potential
approximation in which ${Z(\phi ,t)\equiv }z$ is a constant and ${V(\phi ,t)}
$ a simple function which, sometimes, we conveniently represent as a sum of
monomials of $\phi $ as, e.g., in the case of the Z$_{2}$ symmetry: 
\begin{equation}
{V(\phi ,t)=}\sum_{k=1}^{\infty }u_{2k}\left( t\right) \phi ^{2k}
\label{poly}
\end{equation}

In this approximation~\cite{3480} the ERGE reduces to a differential
equation for ${V(\phi ,t)}$ which, with the Wilson version (\ref{WERGE}),
reads:

\begin{equation}
\dot{V}=V^{\prime \prime }-\left( V^{\prime }\right) ^{2}+\left( 1-\frac{d}{2%
}\right) \phi V^{\prime }+dV  \label{WilLPA}
\end{equation}

In the approximation, the parameter $\eta $ must take on the value $0$.

\underline{Interest}: The local potential approximation allows to consider all powers of $\phi $
on the same footing. It provides us with an excellent textbook example to
illustrate (and to qualitatively investigate) the nonperturbative aspects of
the Wilson theory\footnote{%
Exact results are accessible within this approximation. Wegner and Houghton~%
\cite{414} have shown that the limit $N\rightarrow \infty $ of their ERGE in
the $O\left( N\right) $-symmetric case is identical to the limit $N=\infty $
of the local potential approximation. Now the limit $N=\infty $ corresponds
to the spherical model which is exactly solvable.}. The only lacking
features are related to phenomena highly correlated to the non local parts
neglected in the approximation. For example in two dimensions ($d=2$), where 
$\eta =\frac{1}{4}$ is not particularly small, the local potential
approximation is unable to display the expected fixed point structure.~\cite
{176} Otherwise, when $\eta $ is small (especially for $d=4$ and $d=3$), one
expects the approximation to be qualitatively correct on all aspects of the
RG theory. Especially on the number of existing nontrivial fixed points.

\underline{Fixed points}: The fixed points are solutions of the equation $\dot{V}=0$. This is a non
linear second order differential equation. A nontrivial solution is
parametrized by two arbitrary constants. By imposing, e.g. $V^{*\prime
}(0)=0 $ for an even function of $\phi $, we are left with a one-parameter
family of (nontrivial) solutions to the differential equation. However, all
but a finite number of the solutions in the family are singular at some $%
\phi _{c}$. By requiring the physical fixed point to be defined for all $%
\phi $ then the acceptable fixed points (if they exist) may all be found by
adjusting one parameter in $V(\phi )$.

\underline{Critical exponents}: The best way to calculate the critical exponents is to linearize the flow
equation in the vicinity of the fixed point and to look at the eigenvalue
problem. One obtains a linear second order differential equation. For
example with the Wilson version (\ref{WERGE}), setting 
\[
V(\phi ,t)=V^{*}+\;e^{\lambda t}v(\phi ) 
\]
the eigenvalue equation reads: 
\begin{equation}
v^{\prime \prime }+\left[ \left( 1-\frac{d}{2}\right) \phi -2V^{*\prime
}\right] v^{\prime }+\left( d-\lambda \right) v=0  \label{EigenWil}
\end{equation}

Again one expects solutions to this equation {\sl labelled by two
parameters, however by linearity one can choose }$v(0)=1$ (arbitrary
normalization of the eigenvectors) {\sl and by symmetry} $v^{\prime }(0)=0$. 
{\sl Thus the solutions are unique, given }$\lambda ${\sl . Now for large }$%
\phi ${\sl , }$v(\phi )${\sl \ is generically a superposition of }$v_{1}\sim
\phi ^{2(d-\lambda )/(d+2)}$ and of $v_{2}\sim \exp \left( \frac{d+2}{4}\phi
^{2}\right) $. {\sl Requiring zero coefficient for the latter restricts the
allowed values of }$\lambda ${\sl \ to a discret set''.~}\cite{3357}

The reason for which the exponential must be eliminated is precisely the
necessity of having a quantized set of eigenvalues. This necessity is
related to the notion of renormalizability or, equivalently, of
self-similarity which states that the effect of the infinite number of
degrees of freedom involved in a field theory are all accounted for by means
of a (small) discrete set of renormalized parameters (see part \ref{sec: II}%
). This precise detail is important with respect to the existence of
nonpolynomial solutions giving rise to relevant directions for the Gaussian
fixed point.~\cite{3493} This would make the scalar field theory in four
dimensions asymptoytically free. It exists recent studies of these new
relevant directions,~\cite{3889} however the set of relevant directions is a
continuum and this is not in agreement~\cite{3817,4595}  with the standards of the
renormalization theory (see part \ref{sec: II}).

\underline{Next-to-leading order in the derivative expansion}: There, $Z\left( \phi ,t\right) $ is no longer a constant and a system of two
coupled equations for $V$ and $Z$ takes place. We do not write down them
here, we simply aim at mentioning the breaking of the reparametrization
invariance. This means that instead of having a line of equivalent fixed
points generated by the change of normalization of the field, one gets a
line of non equivalent fixed points each being associated with a different
value of the field-rescaling parameter $\eta $. However a residual effect of
the invariance allows to determine an optimal value\footnote{%
See also J. Comellas. \cite{SchemeDep}} of $\eta $.~\cite{4420,4468}

The derivative expansion generally breaks the reparametrization invariance
except with two regularization schemes:

\begin{itemize}
\item  the sharp cutoff scheme, but the derivative expansion is delicate to
define in that case.~\cite{3550}

\item  the pure power law cutoff function of the form:~\cite{3357,2520}
\end{itemize}

\[
\tilde{C}(p^{2})=p^{2k} 
\]

In that case a unique value of $\eta $ is found and also the eigenvalue $0$
corresponding to the redundant operator ${\cal O}_{1}$ mentioned previously.

\underline{Convergence of the derivative expansion}: One does not know whether the expansion converges or not.~\cite{4326} It
could be Borel summable however.~\cite{4462}

\underline{Conclusion}: Supplementary works on this expansion would be welcome:

\begin{itemize}
\item  Calculation of higher orders

\item  Studies on its convergent or divergent nature

\item  Looking for new kinds of approximations applied to the ERGE (see,
e.g. Golner~\cite{3912}).
\end{itemize}

To those who would have some doubts about the interest of pursuing those
studies, we recommend to look at the rich variety of calculations
that may be done within the nonperturbative Wilson's approach for example in
the papers by Wetterich and coworkers.~\cite{4700}

\section{Wilson's continuum limit (or nonperturbative renormalizability)%
\label{sec: II}}

The expression ``continuum limit'' comes from the consideration of a field
theory on a lattice for which one would like to make the lattice spacing $a$
vanish ($a\rightarrow 0$). In fact, the Wilson theory allows us to consider
special kinds of actions which already display the fantastic property of
evolving (under the effect of a RG transformation) as if the lattice spacing
was actually zero although it is not. These actions are called ``improved
actions''~\cite{2827,2058} or ``perfect actions''~\cite{4283} according to
the method used to construct them\footnote{%
Respectively perturbatively (improved) and nonperturbatively (perfect).}.

Actually, the continuum limit of a lattice field theory leads to a
nonperturbative equivalent expression of the notion of renormalizability%
\footnote{%
A notion inherited from perturbation theory.}. This relies essentially upon
the following correspondence:

\begin{itemize}
\item  in one hand (perturbation theory), the renormalizability states that
one may push the initial cutoff $\Lambda _{0}$ ($\sim \frac{1}{a}$) to
infinity while defining only a finite (usually small) number of renormalized
parameters;

\item  in the other hand (Wilson's theory), in the same limit ($\Lambda
_{0}\rightarrow \infty $), the nature of the renormalized parameters and
their number are given by the finite set (usually small) of relevant
directions of a fixed point.
\end{itemize}

But, be careful, the expression: ``{\em the relevant parameters are the
renormalized parameters}'' is incomplete (and thus incorrect). This is
because there are two {\em inseparable} aspects in the definition of a
renormalized parameter:

\begin{enumerate}
\item  its nature: e.g., is it a $\phi ^{4}$-like or a $\phi ^{6}$-like
coupling or something else?

\item  the RG flow which runs along a renormalized trajectory (RT) emerging
from the fixed point. This flow is usually expressed via a $\beta $-function
expressing a momentum dependence of the renormalized coupling (the nature of
which is already specified).
\end{enumerate}

A relevant parameter has the same nature as the renormalized parameter but,
because the notion of relevance is attached to a linearization of the RG
transformation in the vicinity of a fixed point,~\cite{2835} the flow
carried by a relevant parameter runs locally only {\em tangentially} to the
trajectory of actual interest. Consequently, the second characteristics (the
RG flow along the RT) is not correctly expressed by the relevant character
of a parameter.

It is the aim of the following considerations to illustrate and discuss
these issues.

\underline{Discussion of a traditional figure}: Since the famous review by Wilson and Kogut,~\cite{440} one invariably
illustrates the notion of continuum limit with the once unstable fixed point
configuration giving rise to a {\em purely massive} renormalized field
theory.~\cite{432} But it is not the best candidate\footnote{%
In the purely massive case, the mass may be, at the same time, a scale of
reference and a scale-dependent (renormalized) parameter. This fact obscures
the (momentum) scale dependence.} for an illustration of the most important
aspect of the RG in field theory, namely, the momentum-scale dependence of
the renormalized parameters (i.e., in the continuum limit).~\cite{3554} Let
us first discuss briefly this traditional figure (see figs. \ref{fig1} and 
\ref{fig2}).

In the Wilson space ${\cal W}$ (of infinite dimension $D_{{\cal W}}$) of the
action-parameters, there is the critical submanifold ${\cal W}_{c}$
of dimension $D_{{\cal W}}-1$ in which, when $%
d=3$, lies the Wilson-Fisher fixed point (with only one direction of
infrared instability). If we initialize a bare action in ${\cal W}_{c}$,
then the RG transformation carries the running action toward the fixed
point. If the bare action is chosen very close to ${\cal W}_{c}$, then the
renormalization trajectory passes close to the fixed point and finally goes
away from ${\cal W}_{c}$ essentially along a limiting trajectory which
ideally emerges from ${\cal W}_{c}$ at the fixed point. This trajectory is a
RT because if we formally associate a point of it to a finite and fixed
momentum scale $\Lambda $ then, since an infinite time is required to go out
from the fixed point, the initial cutoff $\Lambda _{0}$ (assumed to be
associated to the fixed point) appears to be infinite with respect to $%
\Lambda $. Then it is an obvious matter to show that the relevant parameter
(presently a mass) takes on a finite value at this point and that,
consequently, a finite (renormalized) mass-parameter exists in the limit $%
\Lambda _{0}\rightarrow \infty $.

This issue, which specifies the nature of the renormalized parameters in the
continuum limit, clearly illustrates the crucial aspect of the
renormalizability: namely the small number of the renormalized parameters
finally involved\footnote{%
The continuum of relevant directions of the Gaussian fixed point found in
four dimensions with nonpolynomial forms of the action~\cite{3493}
contradicts~\cite{3817,4595} this aspect of the RGT (the fixed point then possesses an
infinite number of unstable directions).}.

The second issue (the specification of the RG flow) is the main content of
the RGT. The RG flow in the continuum is precisely (obviously) the flow
which runs along the RT. Ideally, to get it in the exact RGT, we could
initialize the action at a point lying {\em right on} the RT (perfect
action). Although the corresponding cutoff is $\Lambda _{0}$, the perfect
action is ``renormalized'' (the continuum limit is reached) in the sense that the
subsequent evolution under a RG transformation is unique (for a
given fixed point).

Hence, the continuum limit could be
obtained by imposing a perfect action as initial condition to the ERGE. But,
since the RT is entirely plunged in a space of infinite dimension, an
infinite number of action-parameters must be specified. Because this cannot
be realized in practice, one has recourse either to a truncation
(approximate perfect action) or to a limiting process consisting in a fine
tuning of the bare action somewhere\footnote{%
Only one action-parameter (e.g., a bare mass) has to be fine tuned in the
purely massive case.} close to ${\cal W}_{c}$.

In perturbation theory an initial complicated action is never considered
explicitly. On the contrary, one systematically insists on the fact that the
renormalized action remains simple (i.e., does not involve an infinite
number of parameters). Hence, the latter mentioned facet of the Wilson
theory seems to contradict the notion of renormalizability which we were
particularly attached to. 

Actually, in the expression of the RG flow in the
``continuum'', we distinguish two things:

\begin{description}
\item[a)]  the RG flow itself characterized by its instantaneous speed. It
may be expressed with a small number of flowing parameters\footnote{%
Hence the notion of self similarity: the system looks like the same at any
momentum scale of reference.} the nature of which is that of the relevant
parameters at the fixed point. The scale dependence then expresses under a
differential form ($\beta $-functions). It is that aspect of the
renormalization of field theory which is privilegiated in the perturbative
approach.

\item[b)]  the initialization of the RG flow. It requires the specification
of an infinite number of conditions on the action. This aspect of the
renormalization of field theory is nonperturbative by nature. We refer to it
as the functional form of the scale dependence in the continuum limit.
\end{description}

A legitimate question then arises. Since in constructing perturbatively the
renormalized theory we never refer explicitly to a fixed point\footnote{%
Except implicitly to the Gaussian fixed point.} --- and a fortiori to the
notion of relevance ---, and also we never consider any perfect action%
\footnote{%
Except in the Symanzik improvement program~\cite{2827} for lattice field
theory.}, how the right RG flow has been chosen allowing us to get ``the
best'' estimates of the 3-$d$ critical exponents even though these
quantities are highly nonperturbative in nature?

The answer is: by picking out a particular RG flow which turns out to be the
slowest\footnote{%
Asymptotically close to the Gaussian fixed point, this corresponds to having
chosen the relevant flow.} in ${\cal W}$. Let us illustrate this with the
massless $\phi _{3}^{4}$-theory.

\underline{The 3-$d$ massless field theory}: A massless theory is defined in the critical submanifold ${\cal W}_{c}$.
Because the renormalization of field theory gives rise to a scale
dependence, the massless theory of interest is not defined at the fixed
point. Of course, a fixed point theory is scale invariant and could
correspond to a theory without mass (since a mass would break the scale
invariance). But, despite this marvellous and fascinating property of scale
invariance, at a fixed point everything is fixed, there is no scale
dependence and in fact, there is nothing interesting to describe.

The 3-$d$ massless theory which we are interested in is scale dependent and
it is defined by reference to a RG flow running in ${\cal W}_{c}$ and
emerging from the Gaussian fixed point\footnote{%
One usually says that the theory is defined ``at'' the Gaussian fixed point
but it is not an actual fixed-point theory defined at the fixed point.}.

Fig. \ref{fig3} displays the same fixed point configuration as fig. \ref
{fig2} but with a larger view which includes the Gaussian fixed point. We
can see a RT which emerges from the Gaussian FP. It is the RT on which is
defined the continuum limit of the massless field theory for $d=3$.

Fig. \ref{fig4} shows exclusively the (projected) RG flows in the critical
submanifold ${\cal W}_{c}$. We observe critical flows reaching the
Wilson-Fisher fixed point. To get these critical trajectories we must
adjust, for each of them, one parameter of the initial action (critical
temperature).

The RT for the massless field theory interpolates between the Gaussian and
the Wilson-Fisher fixed points. It is an (infrared) attractive one-dimensional
submanifold. It is like a large river into which the ordinary Wilson flows
run. It corresponds to the slowest flow in the Wilson space.~\cite{3554}

Fig. \ref{fig4} is static: it only shows RG trajectories. Fig. \ref{fig5} is
a dynamic picture of the flows showing the evolution of the $\phi ^{4}$%
-coupling $u_{4}\left( t\right) $ along the critical (massless) RT. We
obtain this way an image of the functional forms of the scale dependence in
the continuum which differ only by one initial condition. If we require that
each flow reaches the same value of $u_{4}$ (the same point in ${\cal W}$)
at the same time, then we obtain an unique evolution except for some
marginal ``finite cutoff'' effects corresponding to those trajectories which
were initialized far from the Gaussian fixed point.

Expressed under the differential form, we get a unique $\beta $-function: $%
\beta \left( u_{4}\right) =\Lambda ~du_{4}/d\Lambda $ (see fig. \ref{fig6}).
The $\beta $-function so obtained expresses the uniqueness of the slowest
flow along the RT which interpolates between the Gaussian and Wilson-Fisher
fixed point (in the critical submanifilod ${\cal W}_{c}$). It is precisely
that flow which has been picked out by the perturbative approach while
applying the subtraction procedure required to make finite, order by order,
the perturbative series of the $\phi _{4}^{4}$-theory in the limit $\Lambda
_{0}\rightarrow \infty $. So defined as carrier of the differential RG flow
running along the RT, $u_{4}\left( t\right) $ is like the usual renormalized 
$\phi ^{4}$-coupling of perturbation theory\footnote{%
Notice the freedom in the choice of the renormalized parameter
(arbitrariness of the renormalization scheme). Indeed, it is only an
intermediary in the expression of the essential RG flow of the action
running along the RT. We could have chosen a parameter different from $u_{4}$
provided that asymptotically close to the Gaussian fixed point it behaves
like the $\phi ^{4}$-coupling which coincides with the relevant direction at
this fixed point as shown in fig. \ref{fig4}.}.

Once the ideal flow has been selected near the Gaussian fixed point, it
remains to calculate the perturbative series and to Borel-resum them up to
the Wilson-Fisher fixed point in order to estimate the critical exponents.~%
\cite{283}

Now this is not the whole story, because, when $d=4$, there are ultraviolet
renormalon singularities in the perturbative series which prevent them from
being Borel summable.~\cite{789}

In four dimensions, the Wilson-Fisher fixed point disappears and all the
critical RG flows run toward the Gaussian fixed point (see fig. \ref{fig7}).

We see that, if the ideal slowest flow is underlying (all the trajectories
are attracted to an infrared stable submanifold of dimension one), it cannot
be exactly reached by the RG flows because there is no once-unstable fixed
point in ${\cal W}_{c}$ which could canalize all the degrees of freedom of
the theory along an unique ideal (perfect) trajectory yielding a complete
momentum scale dependence in the range $\left] 0,\infty \right[ $.
Consequently, the Wilson trajectories {\em approach} the ideal trajectory in
the infrared regime but never reach it exactly and there remain ambiguities
between the actual and the ideal trajectories especially in the ultraviolet
regime (i.e., at least for the momentum-scale of reference larger to some
finite $\Lambda _{\max }$). The only possibility of reaching a pseudo ``continuum
limit'' (i.e., of reaching the ideal, but incomplete, slowest RG flow) would be to explicitly
write down the perfect action which associates a point of the underlying
ideal trajectory to $\Lambda $ (initial condition on the functional form of
the momentum scale dependence). However an infinity of conditions are
required\footnote{%
Notice that when a fixed point exists only a few numbers of
action-parameters have to be fine tuned ``at'' the fixed point to reach the
(unique) continuum limit. However an infinity of fine tunings are allowed
giving rise to a family of functional forms of the momentum-scale dependence
(by varying the initial condition).} and moreover the range of variation of
the momentum-scale would remain finite ($\left] 0,\Lambda _{\max }\right] $) due to
the lack of ultraviolet stable fixed point. Ambiguities are then induced in the process of
defining of the initial (perfect) action in the ultraviolet regime.

Ambiguities of the same nature as
the preceding ones exist also in perturbation theory, they are due to ultraviolet
renormalons.~\cite{789} It has been shown by Parisi~\cite{par7} and then by
Berg\`{e}re and David~\cite{berdav1} that the renormalons could be removed
from the theory by considering an infinite number of composite operators and
this would amount to reintroducing a ``finite cutoff''.~\cite{berdav1} In fact, this removal
is nothing but the construction of the perfect action (initialization of the
functional form of the momentum-scale dependence). Indeed it has been shown
that even in the absence of an ultraviolet stable fixed point, the
perturbative series of the $\phi _{4}^{4}$ field theory are Borel summable~%
\cite{172} provided that the RG flow be initialized somewhere and
identifiable at any smaller cutoff scale\footnote{%
That is to say, provided that a definite RG flow is chosen once and for all.}
(infra-red $\phi _{4}^{4}$ model\footnote{%
The infra-red-$\phi _{4}^{4}$-model is a massless theory with an
ultra-violet cutoff. Its renormalization flows have been also rigorously
studied.~\cite{203}}).

In the continuum limit, the problem of dealing with the infinite number of
degrees of freedom is completely involved in the initial condition of the
ideal scale dependence provided we have first picked out the slowest flow.
If the ultraviolet stable fixed point is lacking then the initial condition is obliged
and introduces ambiguities. Otherwise, all the degrees of freedom are exactly canalized in 
the relevant direction leaving no room for any ambiguity (the obliged initial condition takes place ``at''
the fixed point).

Now the following question arises: Why the $\beta $-function perturbatively
determined would display renormalon singularities since it expresses a
simple flow describable by a single variable while the renormalons are
related to the specification of an obliged initial condition (involving
explicitly the infinity of degrees of freedom of the theory)?

The answer is that in the definition of the $\beta $-function, there are two
aspects:

\begin{enumerate}
\item  the slowest flow itself

\item  the running variable chosen to express this flow
\end{enumerate}

If we choose the renormalized coupling via a renormalization-point condition
then the question of dealing
explicitly with the infinity of degrees of freedom (determination of a
vertex function) is raised and there is a gap between the ideal flow and the variable
chosen. Consequently, there are renormalon singularities in the $\beta $%
-function in four dimensions.

However if we choose a renormalization scheme in such a way as to
exclusively refer to the ideal slowest flow without making any reference to
any vertex function, then the $\beta $-function in four dimensions would
have no renormalon singularities at all and it would be Borel summable. Such
renormalization schemes exist, they are called minimal subtraction schemes.
In such schemes, one only subtracts what is necessary to make the theory
finite when $d=4$, the renormalized coupling is not a vertex function, it is
some intermediate parameter not particularly defined in other respects and the
momentum scale of reference is purely artificial. In doing so one has
exclusively referred to the ideal slowest flow. Hence the $\beta $-function
determined in a minimal subtraction scheme could be Borel summable in four
dimensions while the whole field theory remains trivial due to the
renormalons singularities. This proposal agrees with the fact {\sl ``that it
is possible to define renormalization schemes such that the renormalization
group functions [$\beta (u),\eta (u),\cdots $] do not have ultra-violet
renormalons''.~}\cite{berdav1}

Using the same process as above, we could have also discussed~\cite{3554} 
the singularities at the fixed point in the $\beta$-function 
in three dimensions (first envisaged by Nickel~\cite{nic} and also more recently studied~\cite{3541}) 
and the RG trajectories 
in the sector $u_4<0$.~\cite{4627}

\section*{Figure captions}

\begin{enumerate}
\item  The traditional once unstable fixed point configuration, see also
fig. \ref{fig2}.\label{fig1}

\item  This figure, similar to fig. \ref{fig1}, reproduces the projection
onto a plane of actual RG trajectories obtained from a nonperturbative study
in three dimensions of the ERGE in the local potential approximation.~\cite
{3554} It illustrates the simplest nonperturbative continuum limit in three
dimensions:\ Approach to the purely massive ``renormalized trajectory'' $%
T_{0}$ (dot-dashed curve) by RG trajectories initialized at $u_{4}(0)=3$ and 
$u_{n}(0)=0$ for $n>4$ and $(u_{2}(0)-u_{2}^{c})\rightarrow 0^{+}$ (open
circles) in which $u_{2}^{c}$ corresponds to the adjustment of the initial
action in order to approach the Wilson-Fisher fixed point (black full
circle, W.-F.) as in figure \ref{fig4}. The trajectories drawn correspond to $\log
(u_{2}(0)-u_{2}^{c})=-1,-2,-3,-4,-5,-6$. Arrows indicate the infrared direction.\label{fig2}

\item  Same figure as fig. \ref{fig2} with a larger view including the
Gaussian fixed point. It illustrates the continuum limit of the massive $%
\phi _{3}^{4}$-theory defined ``at'' the twice-unstable Gaussian fixed point
and involving two renormalized parameters (a mass and a coupling constant).
The open circles represent initial actions corresponding to different
``fine'' tunings ``at'' the Gaussian fixed point. To get the massless $\phi _{3}^{4}$%
-theory one must adjust, in addition, one action-parameter (e.g., $u_{2}(0)$ to its
critical value $u_{2}^{c}$) in order to make the action running along the ideal
trajectory (T$_{1}$) which interpolates between the two fixed points
(Gaussian and Wilson-Fisher) indicated by full black circles.\label{fig3}

\item  Projection onto the plane $(u_{4},u_{6})$ of some critical RG
trajectories for $u_{4}(0)>0$ [from a numerical study~\cite{3554} with $d=3$%
]. Full lines represent trajectories on the critical submanifold ${\cal W}_{c}$.
The arrows indicate the directions of the RG flows on the trajectories. The
submanifold T$_{1}$ of one dimension (which the trajectories are attracted to)
which links the Gaussian fixed point to
the Wilson-Fisher fixed point corresponds to the renormalized trajectory on
which is defined the continuum limit of the massless field theory in three
dimensions. The open circles represent initial simple actions. \label{fig4}

\item  Functional momentum-scale dependence displayed by $u_{4}(t)$ along
definite RG flows in ${\cal W}_{{c}}$ (lefthand figure). Open circles
indicate the initial points chosen in ${\cal W}_{{c}}$. Each full curve
provides a determination of the functional momentum-scale dependence. In the
righthand figure we have artificially translated the ``time'' scales $t$
(vertical dashed lines) such that each actual Wilson's flow ``hits'' a given
unique value of $u_{4}$ at the same ``time''. The unique functional flow so
obtained illustrates the underlying unique differential flow of fig. \ref
{fig5}.\label{fig5}

\item  Graphical representation of the differential RG flows along the
submanifold T$_{1}$ of fig. \ref{fig4} projected on the $u_{4}$-axis --i.e. $%
\beta (u_{4})=-du_{4}(t)/dt$. In the infra-red direction, any critical RG
flow exponentially approaches the limiting RT T$_{1}$ along which the flows
coincide with the unique differential-momentum-scale-dependence carried by
the ultimate Wilson flow emerging from the Gaussian fixed point (origin).
The
righthand zero corresponds to the Wilson-Fisher infra-red stable
fixed-point-value~\cite{3554} $u_{4}^{*}=3.27039\cdots $. \label{fig6}

\item  RG trajectories on the critical submanifold ${\cal W}_{c}$ when $d=4$
(projection onto the plane $(u_{4},u_{6})$). Open circles represent initial
simple actions chosen in ${\cal W}_{c}$ (of codimension 1)$.$The
two lines which come from the upper side of the figure are RG trajectories
initialized at $u_{4}(0)=20$ and $u_{4}(0)=40$ respectively. The arrows
indicate the infrared direction. The trajectories are attracted to a
submanifold of dimension one before plunging into the Gaussian fixed point.
This is a pseudo renormalized trajectory (it has no well defined beginning)
and strictly speaking, the continuum limit does not exist due to the lack of
another (nontrivial) fixed point which would allow the scale dependence (of
the renormalized parameter along the RT) to be defined in the whole range of
scale $\left] 0,\infty \right[ $. See text for a discussion (from a
numerical study~\cite{3554}). \label{fig7}
\end{enumerate}

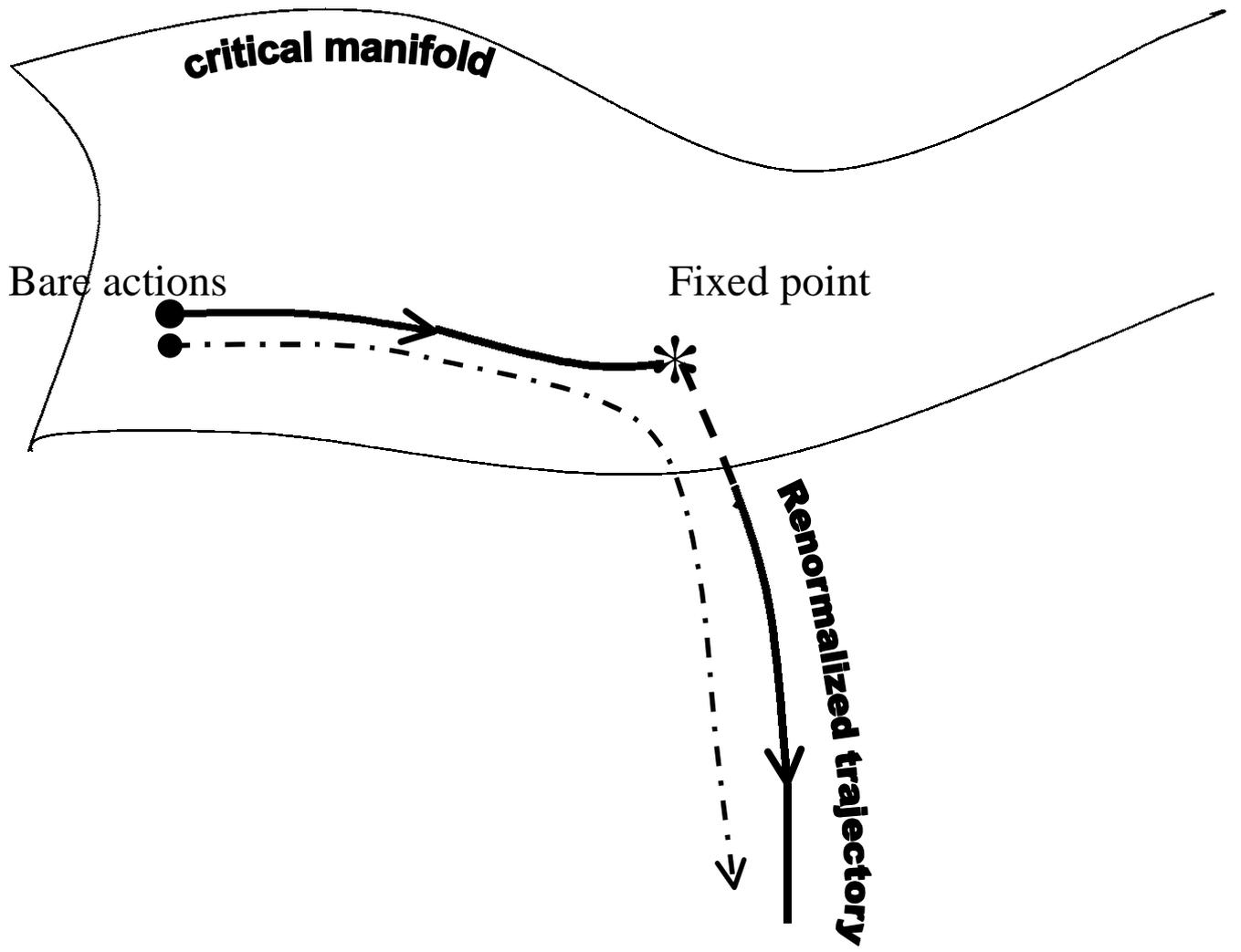
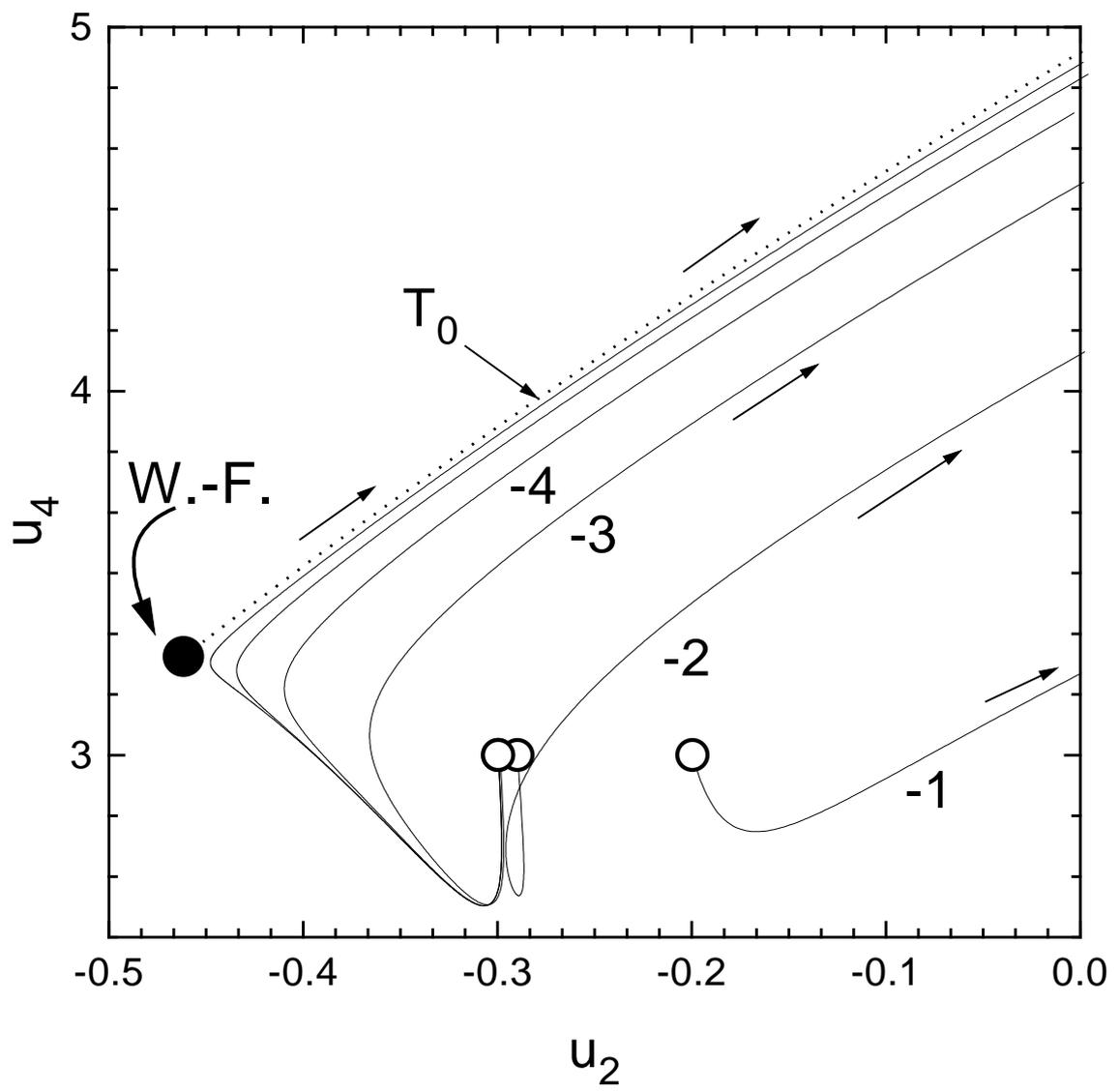
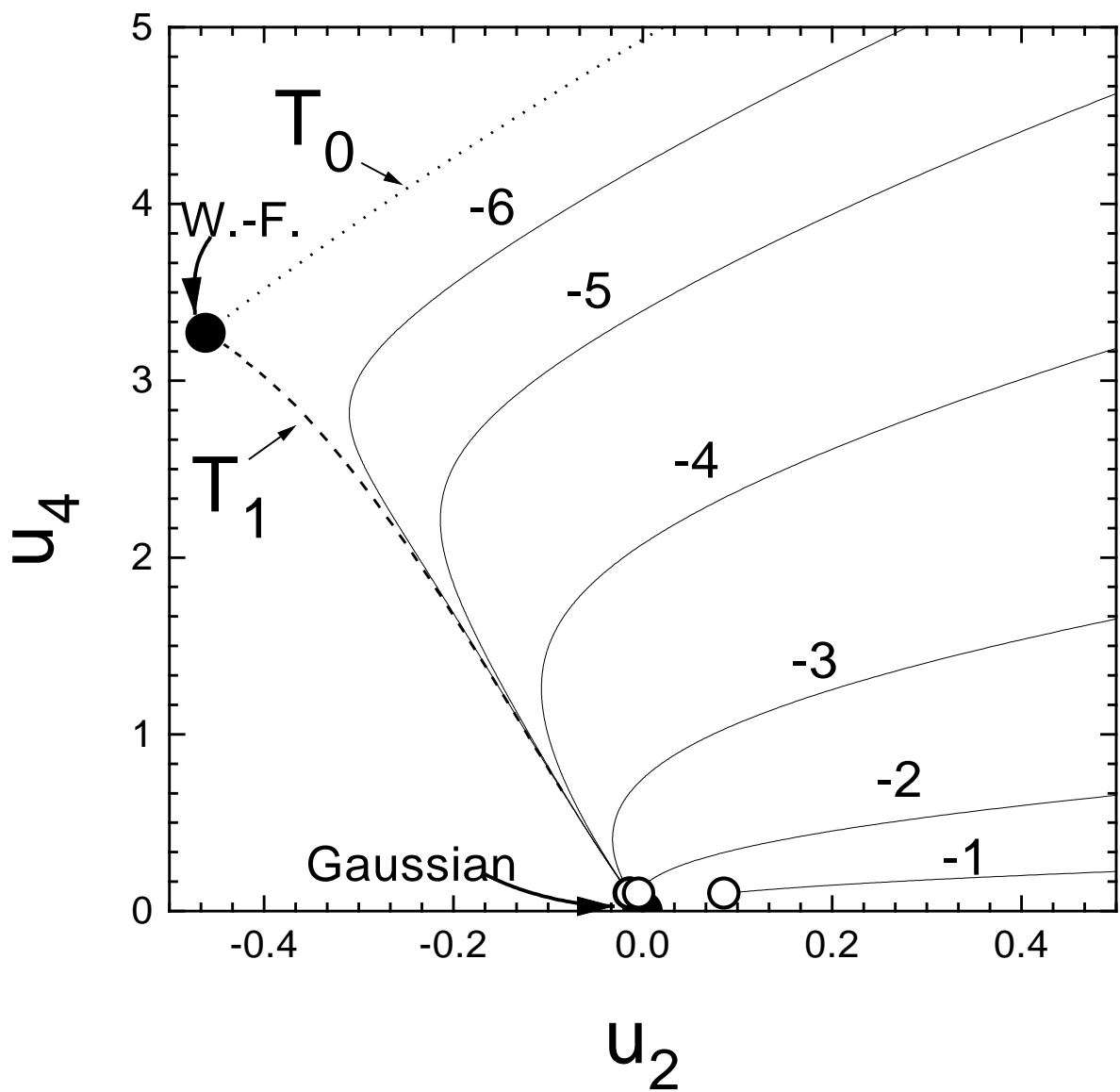
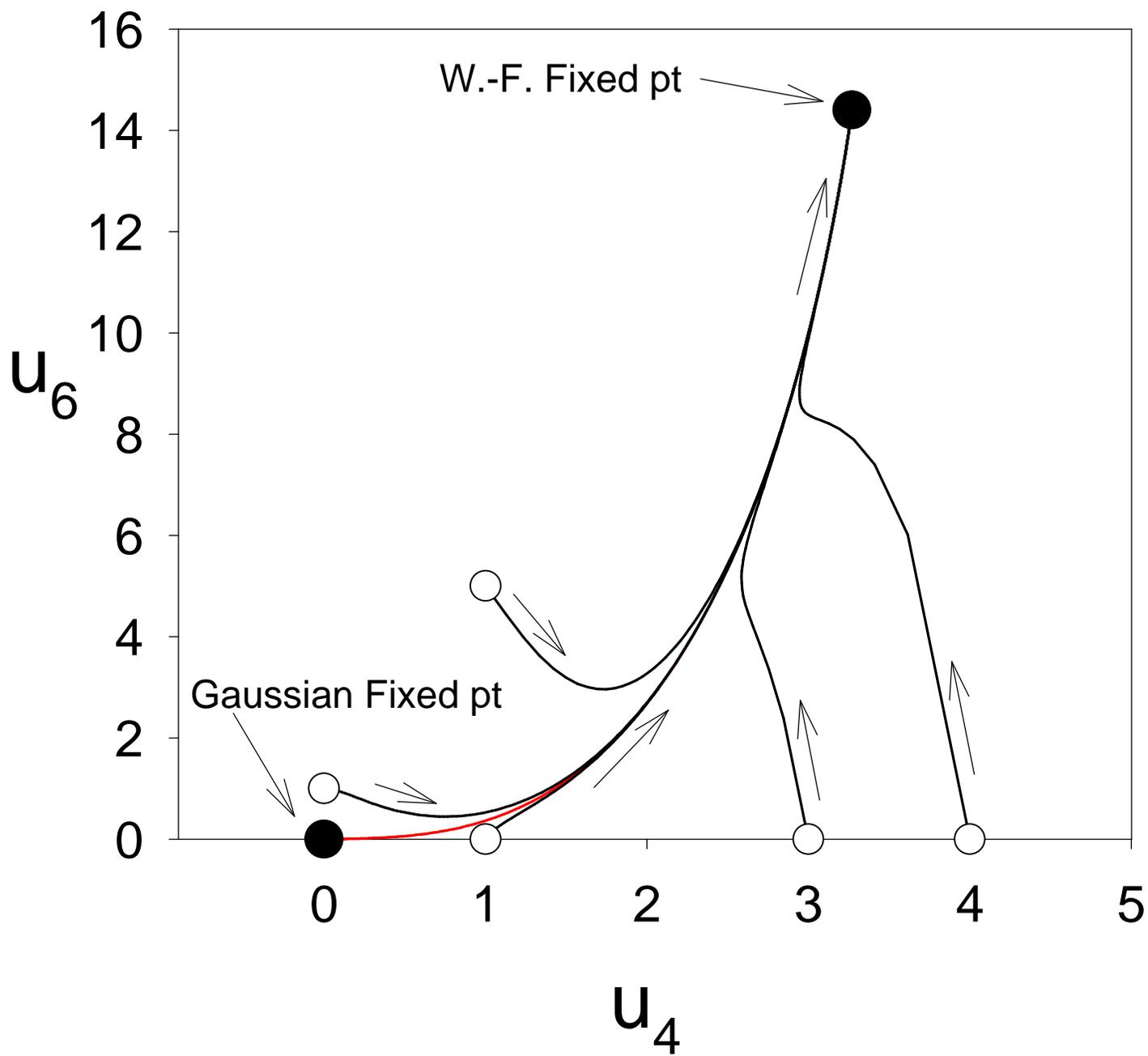
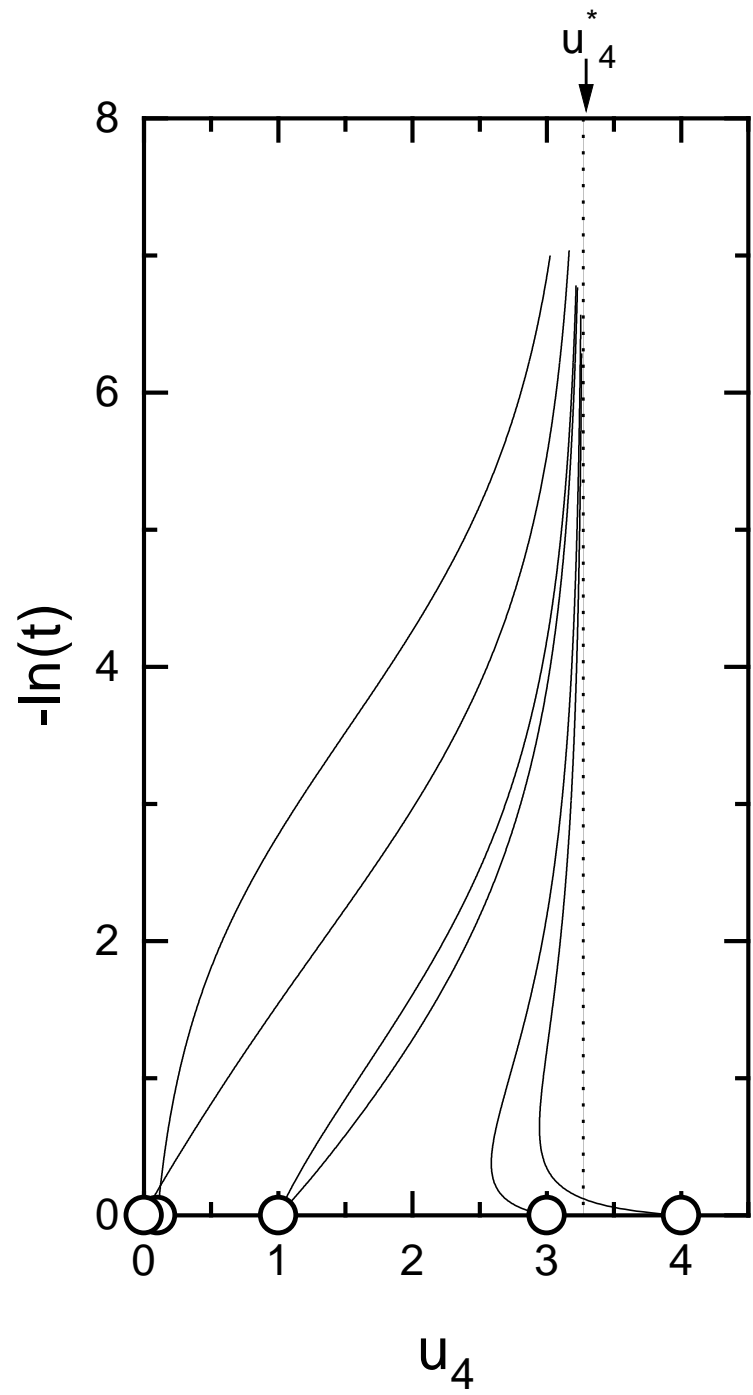
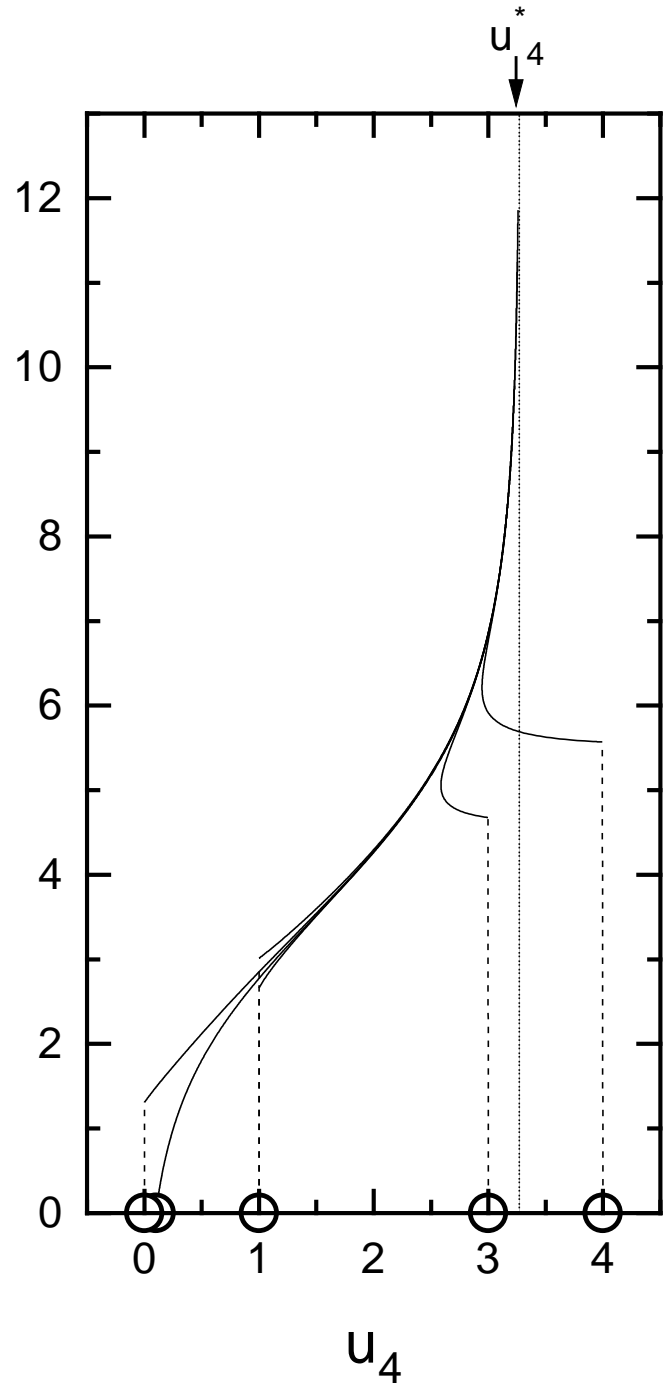
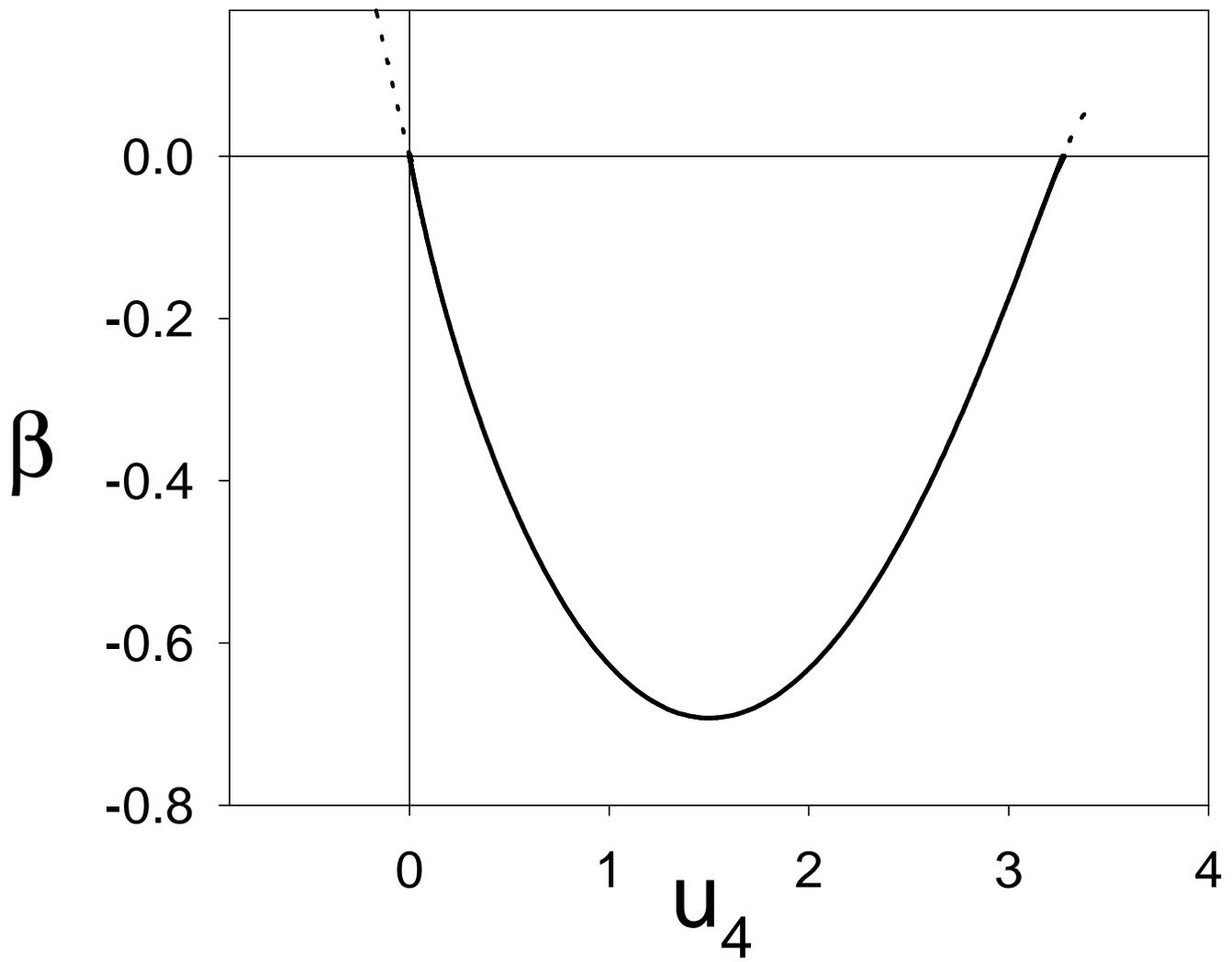
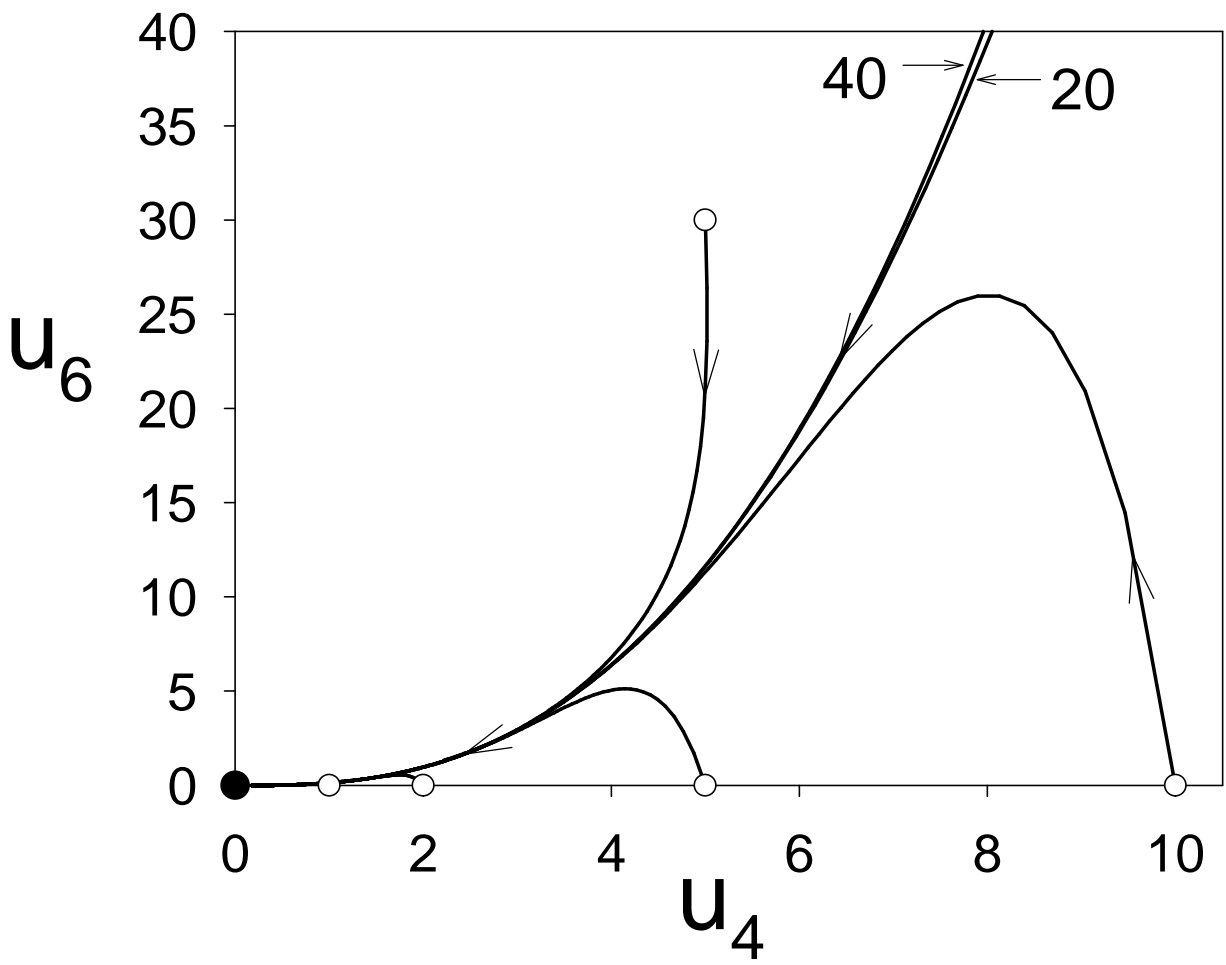
\end{document}